\documentclass[preprint2]{aastex}  

\newcommand{\nc}{\newcommand}
\nc{\bec}{\begin{center}}
\nc{\enc}{\end{center}}
\nc{\beq}{\begin{equation}}
\nc{\enq}{\end{equation}}
\nc{\bei}{\begin{itemize}}
\nc{\eni}{\end{itemize}}
\nc{\bee}{\begin{enumerate}}
\nc{\ene}{\end{enumerate}}
\nc{\namely}{{\it viz.}}
\def\lsun{$L_\odot${}}
\def\micron{\hbox{$\mu$m}}
\def\msun{$M_\odot$}
\def\dep100{$\tau_{\rm 100}$}
\def\dep150{$\tau_{\rm 150}$}
\def\12CO{$^{12}$CO}
\def\13CO{$^{13}$CO}

\begin{document}

\shortauthors{MOOKERJEA et al. }
\shorttitle{FIR study of IRAS 00494+5617 \& 05327-0457}
\slugcomment{ Revised Version \today}


\title{Far Infrared Study of IRAS 00494+5617 \& IRAS 05327-0457}

\author{ B. Mookerjea \altaffilmark{1}, S.K. Ghosh\altaffilmark{1}, T.N. Rengarajan\altaffilmark{1},\\ S.N. Tandon \altaffilmark{2} \& R.P. Verma\altaffilmark{1 } }

\altaffiltext{1}{Tata Institute of Fundamental Research,
             Homi Bhabha Road,
             Mumbai  400 005, India}
\altaffiltext{2}{Inter-University Centre for Astronomy \& Astrophysics,
             Ganeshkhind, Pune 411 007, India}


\begin{abstract}

High angular resolution  far-infrared observations at 143 \& 185 \micron,
using the TIFR 1-m balloon borne telescope, are presented for two Galactic
star forming complexes associated with IRAS 00494+5617 and 05327-0457. The
latter map also reveals the cold dust in OMC-3. The HIRES processed IRAS
maps at 12, 25, 60 \& 100 \micron\ have also been presented for
comparison. Both these regions are illuminated at the edges by high mass
stars with substantial UV flux.The present study is aimed at quantifying
the role of the nearby stars vis-a-vis embedded young stellar objects in
the overall heating of these sources. Based on the FIR observations at 143
\& 185 \micron\ carried out simultaneously with almost identical angular
resolution, reliable dust temperature and optical depth maps have been
generated for the brighter regions of these sources. Radiative transfer
modeling in spherical geometry has been carried out to extract physical
parameters of these sources by considering the observational constraints
like : spectral energy distribution, angular size at different
wavelengths, dust temperature distribution etc. It has been concluded that
for both IRAS 00494+5617 and IRAS 05327-0457, the embedded energy sources
	play the major role in heating them with finite contribution from
	the nearby stars. The best fit model for IRAS 00494+5617 is
	consistent with a simple two phase clump-interclump picture with
	$\sim$ 5\% volume filling factor (of clumps) and a density
	contrast of $\approx$ 80.

{\em Subject Headings}: { IRAS 00494+5617 -- IRAS 05327-0457 
-- Far Infrared Mapping -- Radiative Transfer }
\end{abstract}


\section{Introduction} 

A long term programme of studying the distribution of cold dust ( down to
$\sim$ 15 K) in and around Galactic star forming regions, is in progress
using the 1-m TIFR balloon-borne far-infrared (FIR) telescope. Under this
programme, high angular resolution ($\sim$ 1\arcmin) mapping is carried
out simultaneously in two trans-IRAS FIR wavebands
\citep{ghosh96,bmook98,bmook99a,verma99}. In this paper we present FIR
mapping of the sources IRAS 00494+5617 and 05327-0457 in wavebands
centered around 143 \& 185 \micron. These sources share the similarity
that each is heated by a luminous external source in addition to one or
more possible embedded sources. The simultaneity of observations in the
two wavebands with nearly identical beams is useful in deriving reliable
temperature distributions of the interstellar dust in these regions.

The source IRAS 00494+5617 is  part of the western fragment of the
molecular cloud NGC281, located in the Perseus arm at a distance of 2.2
kpc \citep{ces99}. A compact star cluster containing the multiple star HD
5005 as the brightest star excites an ionization front at the edge of NGC
281W. HD 5005 is an O6 V type star  and the cluster is at an angular
distance of $\sim$ 5\arcmin\ from  IRAS 00494+5617. Observational
evidences in the form of detection of C$^{34}$S (3$\rightarrow$2)
\citep{meg97}, NH$_{3}$ \citep{hen94}, 22 GHz H$_{2}$O masers
\citep{tofani95}, CO emission \citep{carp90} and molecular outflows
\citep{snel90}  together with FIR and millimeter continuum emission,
strongly suggest ongoing star formation activity in this region. Existing
literature on this source suggests that the ongoing star formation
activity has been induced by the compression due to the propagation of the
ionization front (energized by HD 5005) into the molecular cloud
\citep{elme78,meg97}. The gas emission features from this source are
reasonably well studied. There are not as many high angular resolution
observations of the emission from the dust component. Longward of IRAS
wavelengths, the only available observations are the high resolution maps
at 1.3 mm \citep{hen94}.

The source IRAS 05327-0457 is associated with the nebulosity NGC 1977,
located near the northern star in the sword of Orion.  It is bounded on
the south by the northern end of the Orion molecular cloud \citep{kut76}
and is at a distance of 450 pc. The exciting star for the HII region is
the B1 V star HD 37018, also known as 42 Ori and is at an angular distance
of $\sim$ 5\farcm9 from IRAS 05327-0457. Earlier observations of the NGC
1977 region include FIR ($\leq$ 160 \micron), near-infrared and radio
continuum mapping together with extensive molecular line observations
\citep{mak85,kut85}. High resolution molecular line maps \citep{kut85}
indicate an increase in radial velocity  from south to north implying an
expansion of the HII region into the molecular cloud, with a velocity of a
few km/sec.

In the  present study the structural details and dust temperature
distributions obtained from FIR observations have been used to quantify
the relative contributions of the external and internal sources towards
heating these regions. In sections 2 \& 3, the observations and the
results are described. The discussion including the radiation transfer
modeling of these sources is presented in section 4.

\section{Observations}

\subsection{The 143 \& 185 \micron\ maps} The Galactic star forming
regions associated with the sources IRAS 00494+5617 and 05327-0457 were
observed using a two band  far-infrared photometer system at the
Cassegrain focus of the  TIFR 1-m (f/8) balloon-borne telescope. The FIR
telescope was flown from the TIFR Balloon Facility, Hyderabad, in central
India (Latitude = 17\fdg 47 N , Longitude = 78\fdg 57 E )  on 1995
November 12. Details of the telescope and the observational procedure have
been given by \citet{ghosh88}. Additional information specific to this
balloon flight has been presented by \citet{bmook99a}. The photometer
consists of 12 composite silicon bolometers, each having a FOV of 1\farcm6
and arranged in a 3$\times$2 array for each band. The sky was chopped at
10 Hz by wobbling the secondary mirror. The chopper throw was 4\farcm2
along the cross elevation axis. The same region of the sky was viewed
simultaneously in two bands. The effective wavelengths for the two bands
are 143 \micron\ and 185 \micron\ for a greybody spectrum with a
temperature of 36 K and $\lambda ^{-1}$ emissivity. All flux densities
presented in this paper also use the same assumptions regarding
temperature and emissivity. Saturn was observed for absolute flux
calibration as well as for the determination of the instrumental Point
Spread Function (PSF) including the effect of sky chopping. The absolute
flux calibration was done following the method described by
\citet{ghosh88}.

The regions (refer to the Figures~\ref{ngc281} and ~\ref{ngc1977}) around
IRAS 00494+5617 and 05327-0457 were mapped by scanning the sky in
cross-elevation with steps in elevation at the end of each scan line. The
chopped FIR signals were gridded in a two dimensional sky matrix
(elevation \& cross elevation) with a cell size of 0\farcm 3$\times$
0\farcm 3 . The observed chopped signal matrix was deconvolved using an
indigenously developed procedure based on the Maximum Entropy Method (MEM)
similar to that of \citet{gull78} (see \citet{ghosh88} for details). The
FWHM sizes  of the deconvolved maps of the point-like source (Saturn) are
1\farcm 6$\times$1\farcm 9  and 1\farcm 6$\times$1\farcm 8 in the 143 and
185 \micron\ bands respectively. An optical photometer at the Cassegrain
focal plane was used to improve the absolute positional accuracy of the
telescope to $\sim$0\farcm5 \citep{bmook99a}.

\subsection{The HIRES processed IRAS maps}
To supplement our balloon-borne observations, we have used the IRAS survey
data for all four (12, 25, 60 and 100 \micron\ ) bands for the regions of
the sky around  IRAS 00494+5617 and 05327-0457. These data were HIRES
processed \citep{auman90} in the Infrared Processing and Analysis Center
(IPAC\footnote[1]{IPAC is funded by NASA as part of the part of the IRAS
extended mission program under contract to JPL.}, Caltech) for improving
the angular resolutions of the raw maps. These maps have been used  to
obtain flux densities and angular sizes in the four IRAS bands.

\section{Results}

\subsection{IRAS 00494+5617}
Figure~\ref{ngc281} shows the intensity maps of the region around IRAS
00494+5617 in the two TIFR and the four IRAS bands. Table~\ref{ngc281ftab}
presents the coordinates and  flux densities (in a circle of 5\arcmin\
dia) at all the six wavelengths. The main peak at 185 \micron\ is shifted by
$\sim$ 1\arcmin\ compared to the peaks detected at all other wavelengths.
The brightest source at 12 \micron\ (IRAS 00492+5614) is too faint in the
FIR maps and is not the main source of interest.

The 143 and 185 \micron\ maps show more structural details of the region
as compared the 100 \micron\ HIRES map. These maps show extended features
(diffuse component) along the east-west direction at low contour levels.
The presence of a diffuse component is also noticed in all the HIRES maps.
The FIR maps presented here have the shape of a peanut, similar to the
shape at 1.3 mm observed by \citet{hen94}.

In addition to the diffuse component, the 143 \micron\ map shows a
secondary peak (S2) towards the west; this is not seen in the maps of
other bands. The position and flux density (after correcting for the
diffuse emission) of S2 at 143 \micron\ are also presented in
Table~\ref{ngc281ftab}. The limits in the other bands refer to the flux
densities measured in a circle (2\arcmin\ dia) centered at the position of
S2 at 143 \micron. Although  the HIRES maps do not detect S2, the HIRAS
map at 60 \micron\ processed  with the Groningen IRAS Software Telescope
\citep{hen94} with better angular resolution indicates the presence of the
same.

Figure~\ref{ngc281tmptau} presents the contour maps of the temperature
(T(143/185)) and optical depth (\dep150) for this region, generated using
the method described by \citet{bmook99a}. For these maps a dust
emissivity, $\epsilon_{\lambda}$ $\propto$ $\lambda^{-1}$ has been
assumed. The temperatures determined are accurate to within $\pm$ 2 K
between 20 and 50 K and within $\sim$ $\pm$ 5 K between 50 and 75 K. The
structural details in these maps are highly reliable due to (i) the
simultaneity of observations and (ii) conservative processing e.g., 3
pixel by 3 pixel smoothing  of the flux densities done prior to the
determination of the temperature. The temperature and optical depth maps
have been restricted to the regions where the intensities in both the
wavebands are substantially higher ($>$ 10 times) than the measured noise
level. From the temperature map it is seen that there are two regions of
enhancement close to the boundary, the highest temperature being close to
the boundary facing HD 5005. Further, the temperature distribution seems
to be featureless over most of the central region and shows a minimum near
the center. In contrast, the peak of the \dep150\ map coincides with the
peak at 185 \micron\ and the  contours decrease smoothly outwards, as in
the intensity map.

\subsection{IRAS 05327-0457}
Figure~\ref{ngc1977} presents the intensity maps of the region surrounding
IRAS 05327-0457 in the six wavebands considered in this paper. The
positions of the global peak in the 143 \& 185 \micron\ maps agree with
the IRAS Point Source Catalog (PSC) coordinates within the achieved
absolute positional accuracy in the maps. The source is extended along a
direction approximately perpendicular to the line joining the cloud to the
star  42 Ori. The basic features of the intensity distributions at 143
and 185 \micron\ are similar to the 60 and 100 \micron\ HIRES maps as well
as the 158 \micron\ [C II] maps by \citet{howe91}. The emission due to 42
Ori (located to the north of IRAS 05327-0457) is clearly seen in the 12
and 25 \micron\ maps. Table~\ref{ngc1977ftab} presents the position and
flux densities in a circle of 5\arcmin\ dia centered on the global peak in
each of the six wavebands.

In both the  143 and 185 \micron\ maps, a fainter source (P2) is detected
towards the south of IRAS 05327-0457. Positionally this matches very well
with the coldest sub-millimeter source (CSO 10) detected at 350 \micron\
by \citet{lis98}. This is associated with the Orion Molecular Cloud -3 and
has also been detected at 1.3 mm by \citet{chini97}. The dust temperature,
T(143/185) for P2 has been estimated to be  $\sim$ 19 K assuming
$\epsilon_{\lambda}\propto\lambda^{-2}$, the choice of the dust emissivity
index being guided  by \citet{lis98}. This low temperature source is not
detected in any of the HIRES maps. Table~\ref{ngc1977ftab} presents the
position and flux densities of P2.

Figure~\ref{ngc1977tmptau} shows the maps of the dust temperature
(T(143/185)) and optical depth (\dep150) for IRAS 05327-0457, generated
using same methods as for IRAS 00494+5617. In the \dep150\ map, the peak
has a value of 0.02 and is shifted, north-east towards the direction
of the ionizing star. The optical depth decreases monotonically towards
the south-west. Higher temperatures are seen towards the south-eastern and
western edges; rest of the region is seen to have smoothly varying
temperatures, between 25 and 35 K. There is no increase in the dust
temperature near the edge facing 42 Ori. The T(143/185) and the \dep150\
map together suggest the presence of (i) embedded sources of heating at
the positions of temperature enhancements and (ii) a high density shell
facing 42 Ori. 

\section{Discussion}

\subsection{Radiation transfer scheme \& Models explored \label{secrad}}

We have explored the major sources of dust heating based on radiation
transfer modeling of the sources. The radiation transfer equations have
been solved assuming a 2-point boundary condition  for a spherically
symmetric cloud of dust and gas. Based on the boundary conditions, we have
constructed 3 types of models : $model~ A$ -- the cloud is  heated by
centrally embedded sources and  by an external radiation field due to the
average Galactic Interstellar Radiation Field (ISRF) and nearby stars;
$model~ B$ \-- the cloud is heated by internal sources and ISRF only;
$model~ C$ -- there is no embedded source, the cloud is heated only
externally by the ISRF and radiation due to nearby stars. The contribution
of the nearby star in the models $A$ and $C$ has been calculated in the
following way: the geometrically diluted stellar radiation intercepted by
the cloud surface is estimated and it is then smeared out uniformly over
the entire surface of the cloud. This calculated contribution could be a
slight overestimate since the absorption in the intervening medium has
been neglected. For modeling the observed Spectral Energy Distribution
(SED) two types of dust have been considered. The dust properties have
been taken from two sources viz., \citet{draine84} (hereafter DL) and
\citet{mez82} (hereafter MMP). The dust compositions include mixtures of
astronomical silicate and graphite. For the DL type of dust grains, size
averaged properties are used taking a size distribution of $n(a)da \sim
a^{-3.5}da$ \citep{mrn77}, with $a$ ranging between 0.01 and 0.25 \micron.
The average Galactic ISRF used for this problem has been taken from
\citet{mmp83}. The cloud is parameterized by the following physical
quantities: R$_{\rm max}$, the outer size of the cloud, R$_{\rm min}$, the
radius of the inner dust cavity, $\tau_{\rm 100}$, the total radial
optical depth at 100 \micron\ and the radial dust (hence gas) density
distribution ($r^{\beta}$, $\beta$=0, -1, -2). The contributions of ISRF
and the external source (if any) are kept fixed, while the luminosity of
the embedded source, the dust composition, $\tau_{\rm 100}$ value and
radial distribution of the dust density and the physical sizes of the
cloud are varied to obtain a good match to observations. The gas to dust
ratio has been assumed to be 100:1 by mass. Using this scheme a best fit
model matching the observed spectral energy distributions (SED), radial
profiles at selected wavelengths and the radio continuum flux are
obtained. These radiation transfer calculations have been done using a
modified version of the code CSDUST3 \citep{egan88}, which is capable of
treating gas and dust in a self-consistent manner
\citep{bmook99a,bmook99b}.

In the following subsections we explore the tenability of the models $A$,
$B$ and $C$ in the light of available observational constraints for
individual sources.

\subsection{IRAS 00494+5617}

\subsubsection{ The main source } 

The observed SED for the source IRAS 00494+5617 has been constructed using
the flux densities presented in Table~\ref{ngc281ftab} along with flux
densities at 1.3 mm \citep{hen94} and 2.9 mm \citep{walk90}. The total
observed luminosity is 2.4$\times$10$^{4}$ \lsun\ \citep{carp90} and  no
radio continuum emission is detected at 8.4 GHz \citep{tofani95}  from
this cloud. T(143/185) map shows enhanced temperature towards the boundary
closer to HD 5005. In order to explain these, the embedded source in
radiation transfer models $A$ and $B$ is assumed to be a cluster of stars.
In models $A$ and $C$ an intercepted luminosity of
6.9$\times$10$^{3}$\lsun\  from HD 5005 has been considered. For each type
of model it is found that the best fit is obtained for the case of uniform
density and grains of DL type. Figure~\ref{ngc281sed} shows  the observed
and the best fit SEDs. Model $A$ fits the SED at all wavelengths very
well, while  model $B$ shows some mid-infrared excess. Model $C$ fails to
reproduce any of the observed flux densities. The predicted radial
profiles (as obtained by convolving the model profiles with instrumental
PSFs) for the models $A$ and $B$ are almost similar. Table~\ref{proftab}
compares the FWHMs predicted by model $A$ with those observed at 25, 60,
100, 143 and 185 \micron. The observed FWHMs presented here refer to the
FWHMs for the source along its major and the minor axes. We find that
model $A$ not only fits the observed SED better than model $B$ but also
achieves an overall consistency by supporting the role of HD 5005 in
heating the cloud and fitting the observed FWHMs reasonably well.
Table~\ref{tranparm} presents the parameters of the best fit model $A$ for
this source alongwith the uncertainties in the $\tau_{\rm 100}$ and the
embedded luminosity. These uncertainties have been estimated using only
the SED fit and keeping the other parameters fixed. From this radiation
transfer modeling and the T(143/185) map we conclude that the source IRAS
00494+5617 is primarily heated by one or more embedded sources with
non-negligible contribution from HD 5005.

We compare he FWHM predicted from the best fit spherically symmetric model
with the observed FWHM along the minor axis. For IRAS 00494+5617 we find
that at 100, 143 and 185 \micron\ model predictions for FWHMs match the
observed FWHMs (along minor axis) fairly well. However at 25 \micron\
(which traces much hotter dust) the observed FWHM is substantially larger
than the model prediction. This trend is also seen at 60 \micron, though
not to the same extent as at 25 \micron. This can be explained by invoking
spatially distributed heating sources embedded in clumps. It may be noted
that our dust temperature maps also shows the presence of two temperature
enhancements. This clumpiness is also substantiated by high resolution
C$^{18}$O observations by \citet{meg97}, which have revealed that this
western fragment of NGC 281 is composed of three clumps. The best fit
model $A$ obtained above corresponds to a gas density of
2.2$\times$10$^{4}$ cm$^{-3}$ ( for a gas to dust density of 100:1 by
mass) and a total mass of 1890\msun. \citet{meg97} obtain 1080\msun\ as
the lower limit for the total mass in the three clumps. The residual mass
(810 \msun) as predicted by the model can be present in the form of the
inter-clump medium. The projected radii (taking the distance to the source
to be 2.9 kpc) of the clumps were given by \citet{meg97} assuming the
clumps to be spherical. We assume that all three clumps are spherical and
are confined within the boundary of the spherical model $A$ above. After
scaling the clump sizes to our assumed distance of 2.2 kpc, we find that
the volume filling factor is 4.5 \%. Using the total mass from our model
$A$ and assuming an interclump medium of uniform density we obtain a value
of 4.5$\times$10$^{3}$ cm$^{-3}$  for the number density of the interclump
medium. We also calculate the clump-interclump density contrast to be
$\sim$ 80. The resultant clump density of 3$\times$10$^{5}$ cm$^{-3}$ is
consistent with the detection of C$^{34}$S. Although the estimates
presented here are based on crude approximations, the values for filling
factors of clumps and clump-interclump density contrast obtained here can
be called very typical, while comparing with the measured values of such
parameters for other clumped sources \citep{howe91}.

\subsubsection{Diffuse emission} 

The 143 \micron\ map of this region shows a considerable amount of diffuse
emission particularly towards the west. This emission is to a certain
extent positionally coincident and structurally similar to that  in the
185 \micron\ map. We have looked into the possible energy sources for this
emission. The flux densities (in an aperture of 2\arcmin\ dia centered at
0$^{\rm h}$ 49$^{\rm m}$ 09$^{\rm s}$ +56\arcdeg\ 17\arcmin\ 30\arcsec\
(1950)) at 143 and 185 \micron\ are 236 and 265 Jy respectively. The
bolometric luminosity of the diffuse emission over a diameter of 2\arcmin\
has been estimated to be $\sim$ 870 \lsun. This luminosity cannot be
explained due to heating by either the the average ISRF or the star
cluster HD 5005. We explore the possibility of radiation actually leaking
from the main source (around 00494+5617) and heating the dust in the
neighbourhood. Under this assumption we obtain that about 8\% of the total
luminosity of the embedded source would have to be intercepted by the
region where diffuse emission is detected. The separation between the
region of diffuse emission and IRAS 00494+5617 is such that the solid
angle covered by the diffuse region is adequate to intercept the requisite
amount of radiation. This evidence of luminosity leakage further supports
the presence of clumpiness discussed in Section 4.2.1.

\subsection{IRAS 05327-0457}

The observed SED for the region around IRAS 05327-0457 has been
constructed using the flux densities presented in Table~\ref{ngc1977ftab}
alongwith the IRAS Low Resolution spectrum (LRS) \citep{chen95}. VLA
observations of this source by \citet{kut85} have not detected any radio
continuum emission at 5 GHz and from the detection limit predict the
embedded source to be a ZAMS B2.5 or later type star. Based on their
observations \citet{kut85} have suggested that 42 Ori plays the most
significant role in fueling the FIR emission from this source. However the
T(143/185) map presented here shows no temperature enhancement towards the
boundary illuminated by the star 42 Ori. It rather shows an increase from
the boundary towards the position of IRAS 05327-0457. 

We have applied the radiation transfer models $A$, $B$ and $C$ (see
Section~\ref{secrad}) to IRAS 05327-0457. The embedded source for models
of type  $A$ and $B$ is taken to be a star of type B2. In  models of types
$A$ and $C$ the intercepted contribution of 600\lsun\ from 42 Ori is
considered. Figure~\ref{ngc1977sed} presents the SEDs predicted from the
models $A$, $B$ and $C$ in comparison with the observations. The model
outputs shown are the best fits for each type. As in the case of IRAS
00494+5617 uniform density gives the best fit for all three models. Further the
MMP type grains lead to better fit than the DL type grains. While
the predicted SED from model $A$ matches the observations at all wavelengths
reasonably well, model $B$ predicts fluxes less than the observed values
longward of 60 \micron. Predictions from model $C$ do not the fit the
observations at all. The predicted FWHMs (post convolution) are similar
for models of type $A$ and $B$. Table~\ref{proftab} presents the
comparison of FWHM sizes from model $A$ at different wavelengths
with observations. From model $A$ the predicted radio continuum
emission at 5 GHz is $\sim$14 mJy which is well above the
sensitivity of the VLA observations by \citet{kut85}. The reasons
for this  non-detection  could be : (i) the embedded source is a
cluster of stars (with masses less than that of a B2 star)
with combined luminosity equal to that of a B2 star. 
This is supported by the detection of multiple stars in
the near-infrared \citep{mak85}. They have detected 7 stars having a
total luminosity of 2.3$\times$10$^{3}$\lsun\ (model $A$
total luminosity = 2.8$\times$10$^{3}$\lsun). 
Our temperature map also indicates presence of
two sources, again close to the boundary. (ii) Since the
radio continuum emission has a strong n$_{e}^{2}$
dependence, a smaller gas to dust ratio could explain the
reduced emission. (iii) The radio emission could largely
be attenuated by free-free absorption. The third
possibility proposed would arise, if the gas density
distribution very close to the star (r$<<$R$_{\rm min}$)
is r$^{-2}$, a typical situation for ionized wind.

Of the models presented here, model $A$ explains the observed SED, the
radial profile, the T(143/185) map and the role of 42 Ori in heating the
cloud most satisfactorily. Table~\ref{tranparm} presents the parameters of
this model and the estimated uncertainties in the embedded luminosity and
$\tau_{\rm 100}$. From radiation transfer modeling we conclude that it is
absolutely necessary to have an embedded source to explain the total
emission from IRAS 05327-0457. Clearly 42 Ori cannot be the only heating
source since the luminosity intercepted by IRAS 05327-0457 is $\leq$ 20 \%
of the observed luminosity. Further we have verified that the temperature
profile from radiation transfer calculation using a slab geometry and 42
Ori as the source, does not fit observations.

The observed angular sizes for IRAS 05327-0457 at 100 \micron\ and more
prominently at 60 \micron\ are larger than the model values. As in IRAS
00494+5617 this is again consistent with a distribution of embedded hot
sources as suggested by the T(143/185) map (Figure~\ref{ngc1977tmptau}) as
well. The apparent reduction in the observed FWHMs at 143 and 185 \micron\
as compared to the FWHM at 100 \micron\ could be due to the effect of sky
chopping.

\section{Summary} 

This paper presents simultaneous far-infrared mapping observations at 143
and 185 \micron\ of the regions around IRAS 00494+5617 and IRAS
05327-0457. Both sources have been well resolved at these wavelengths. The
cold dust source  CSO 10 located in OMC-3 \citep{lis98}  has also been
detected in our map of IRAS 05327-0457. HIRES processed IRAS maps in all 4
bands have been used for comparison. Reliable dust temperature
(T(143/185)) and optical depth (\dep150) maps for these sources are
presented. Radiation transfer models have been constructed to explain the
observed SEDs, radial profile and the influence of external heating
agencies for both these sources. These models provide useful physical
parameters for the star forming clouds associated with these sources. For
both IRAS 00494+5617 and 05327-0457 it has been demonstrated  that
although HD 5005 and 42 Ori contribute to their heating respectively, it
is necessary to have dominant contribution from embedded sources to
explain the observed SEDs. Comparison of predicted radial profiles from
models with observations suggest presence of a distribution of embedded
clumps in both sources. This is corroborated by spatially resolved peaks
in the maps of dust temperature. A simplistic model for the clumpiness of
IRAS 00494+5617 has been proposed based on parameters derived from
radiation transfer models and previous molecular line observations.

\acknowledgements
It is a pleasure to thank all members of the Infrared Astronomy Group of
T.I.F.R.  for their support during laboratory tests and balloon flight
campaigns.  All members of the Balloon Group and Control Instrumentation
Group of the TIFR Balloon Facility, Hyderabad, are thanked for their
technical support during the flight. We thank IPAC, Caltech, for providing
us the HIRES processed IRAS products. We thank the anonymous referee for
the suggestions which have improved the quality of the paper.



{}

\figcaption[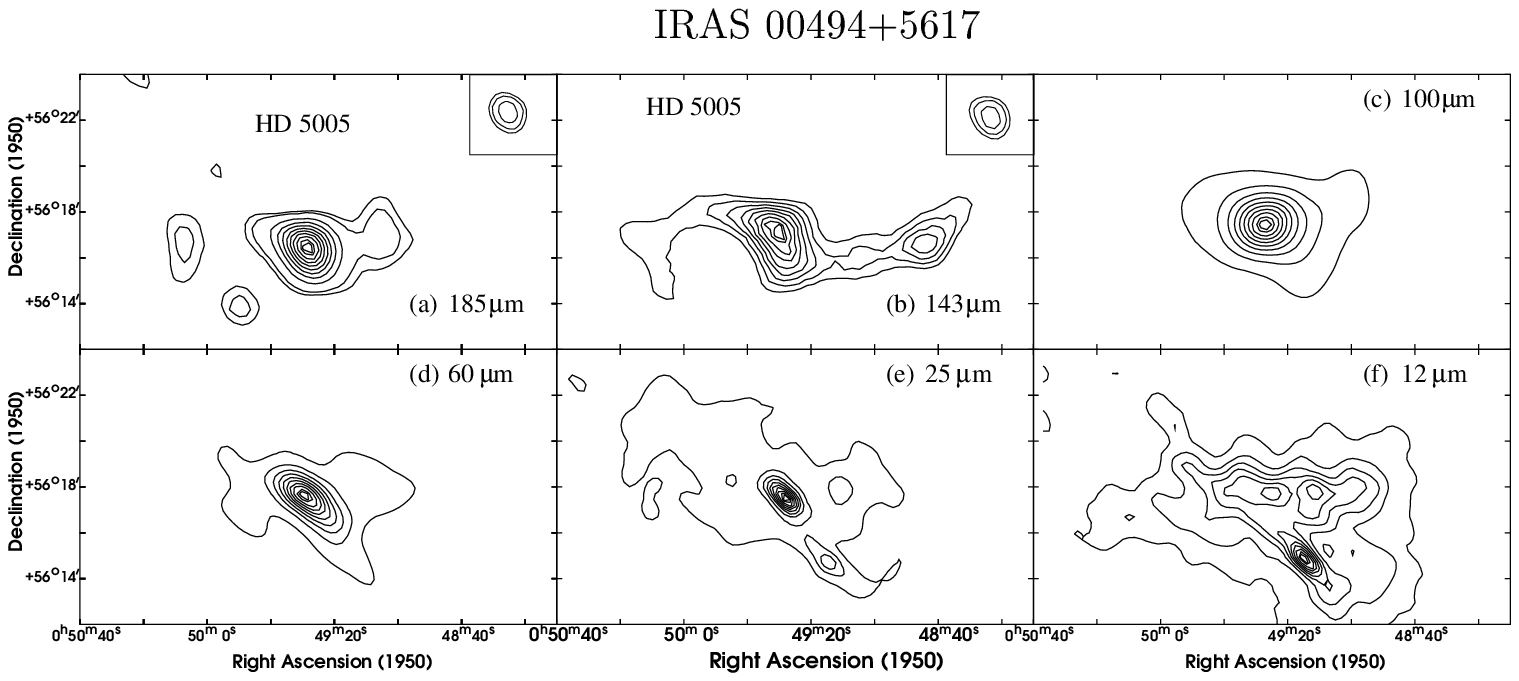] 
{ The intensity maps for the region around IRAS 00494+5617 : (a) at 185
\micron\ with peak = 644 Jy/sq. arcmin  with contour levels =95, 90, 80,
70, 60, 50, 40, 30, 20, 10, 7.5 \% of the peak;  (b) at 143 \micron\ with
peak = 341 Jy/ sq. arcmin. Contour levels in (b) are same as in (a) but
upto 20 \% of the peak. ``$\ast$'' shows the position of HD 5005. The
lowest contour for each map is $\sim$ 5 times the noise level. The insets
show deconvolved images of Saturn in the respective bands, aligned to the
instrumental axes for meaningful comparison. The contours for Saturn
denote 90, 70 and 50 \% of respective peaks. HIRES processed IRAS maps at
(c) 100 \micron, (d) 60 \micron, (e) 25 \micron\ and (f) 12 \micron, with
contours at levels same as in (a) but upto 10\% of the respective peaks.
The peak values in these bands  are 302, 173, 20.9 and 5.6 Jy /sq. arcmin
respectively. The resolutions at 12, 25,  60 and 100 \micron\ are
0\farcm9$\times$0\farcm45, 0\farcm7$\times$0\farcm45,
1\farcm3$\times$0\farcm8 and 2\farcm0$\times$1\farcm6 respectively .
\label{ngc281} }

\figcaption[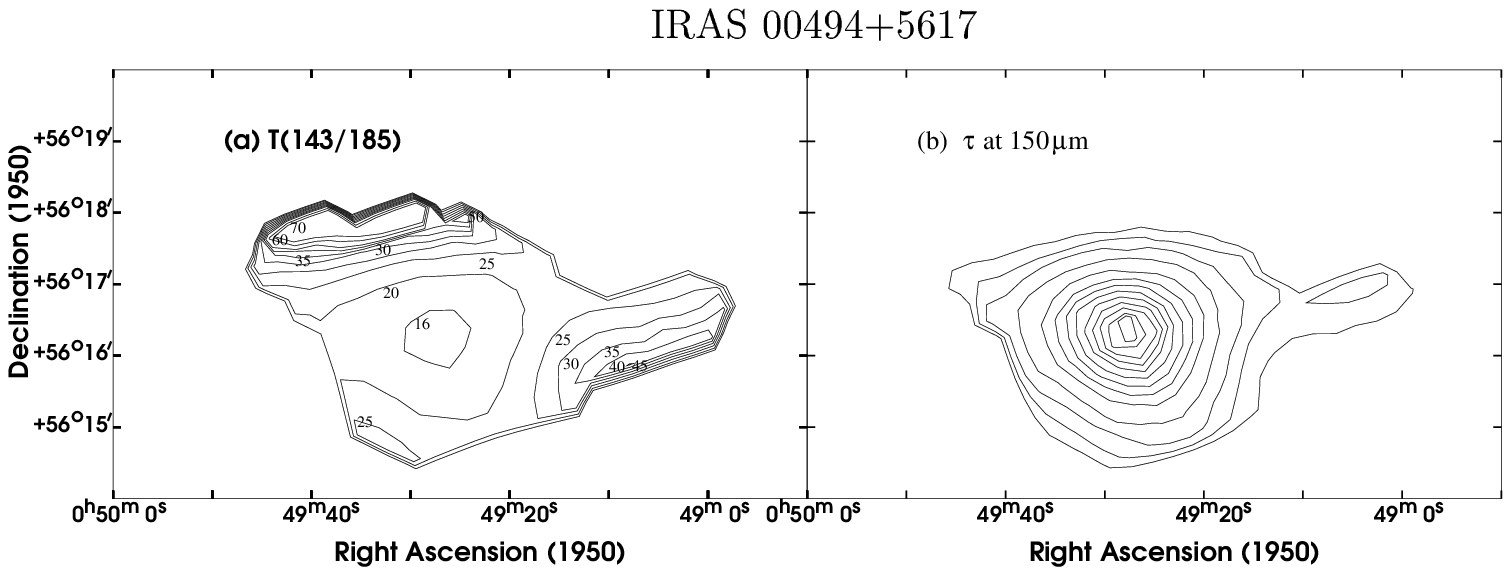]
{ (a) The dust temperature, $T(143/185)$ and (b) optical depth \dep150,
maps respectively for the region near IRAS 00494+5617 for
$\epsilon_{\lambda}\propto\lambda^{-1}$ emissivity law. The contours in
(a) are at 70, 60, 50, 45, 40, 35, 30, 25, 20 \& 16 K. The contours in (b)
are at 95, 90, 80, 70, 60, 50, 40, 30, 20, 10, 5, 2.5, 1\% of the peak
(0.15). 
\label{ngc281tmptau} 
}

\figcaption[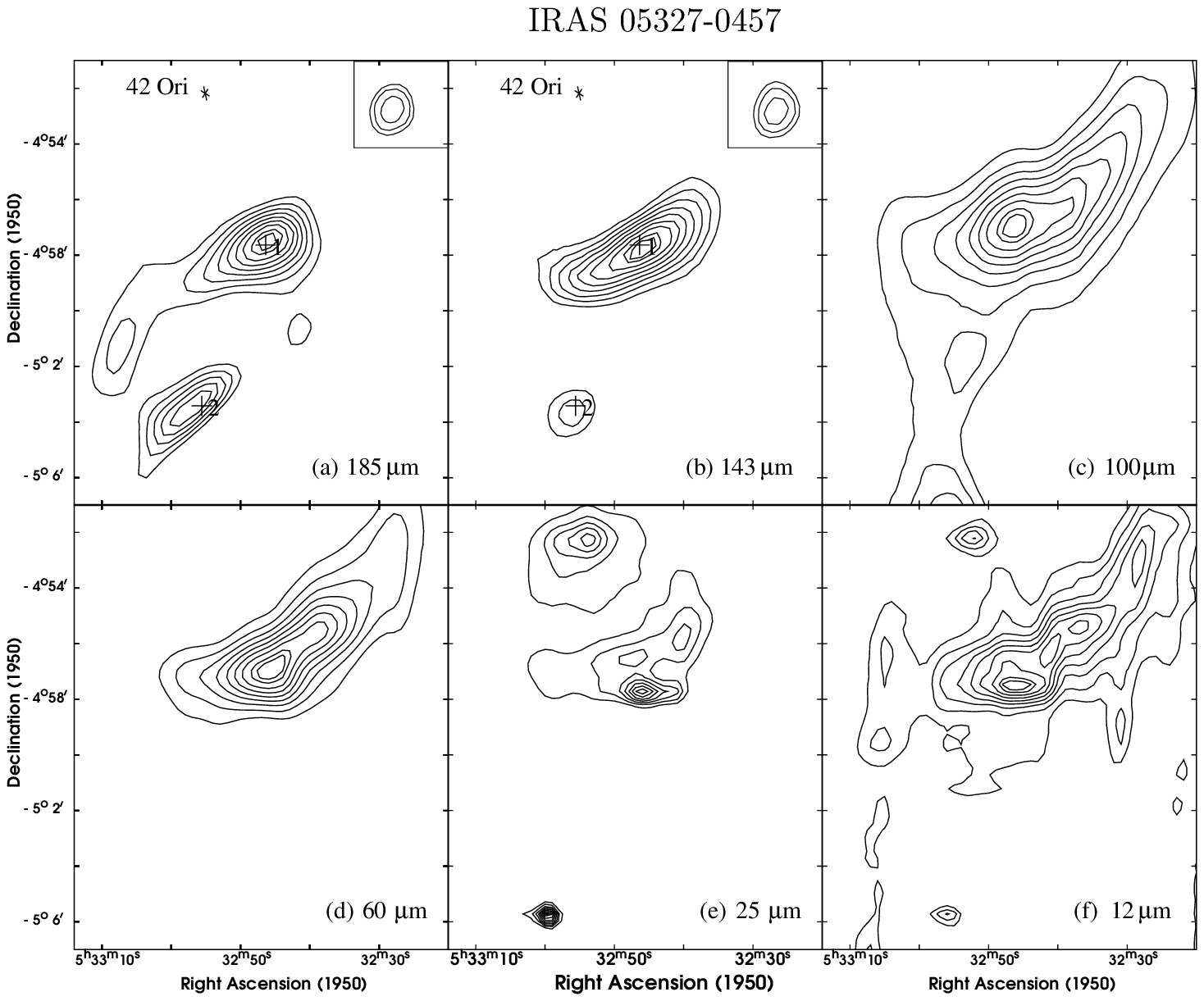]
{ The intensity maps for the region around IRAS 05327-0457 : (a) at 185
\micron\ with peak = 548 Jy/ sq. arcmin and (b) at 143 \micron\ with peak
= 681 Jy/ sq. arcmin.``+1'' shows the IRAS PSC position of the source,
``+2'' shows the position of the source CSO 10 (Lis et al., 1998) and
``$\ast$'' shows the position of 42 Ori.The lowest contour is $\sim$ 5
times the noise level. Insets are same as in Fig. 1(a) and (b). HIRES
processed IRAS maps at (c) 100 \micron, (d) 60 \micron, (e) 25 \micron\
and (f) 12 \micron. with contours at levels same as the top 9 levels of
Fig. 1 (a). The peak values in these bands  are 802, 688, 83.9 and
25.5 Jy /sq. arcmin respectively. Contours levels at all wavelengths are
same fractions of the respective peaks as in Figure 1(b). The angular
resolution at 12, 25, 60 and 100 \micron\ are 1\farcm0$\times$0\farcm3,
1\farcm1$\times$0\farcm6, 1\farcm6$\times$1\farcm1 and
2\farcm2$\times$2\farcm0 respectively. 
\label{ngc1977} 
}

\figcaption[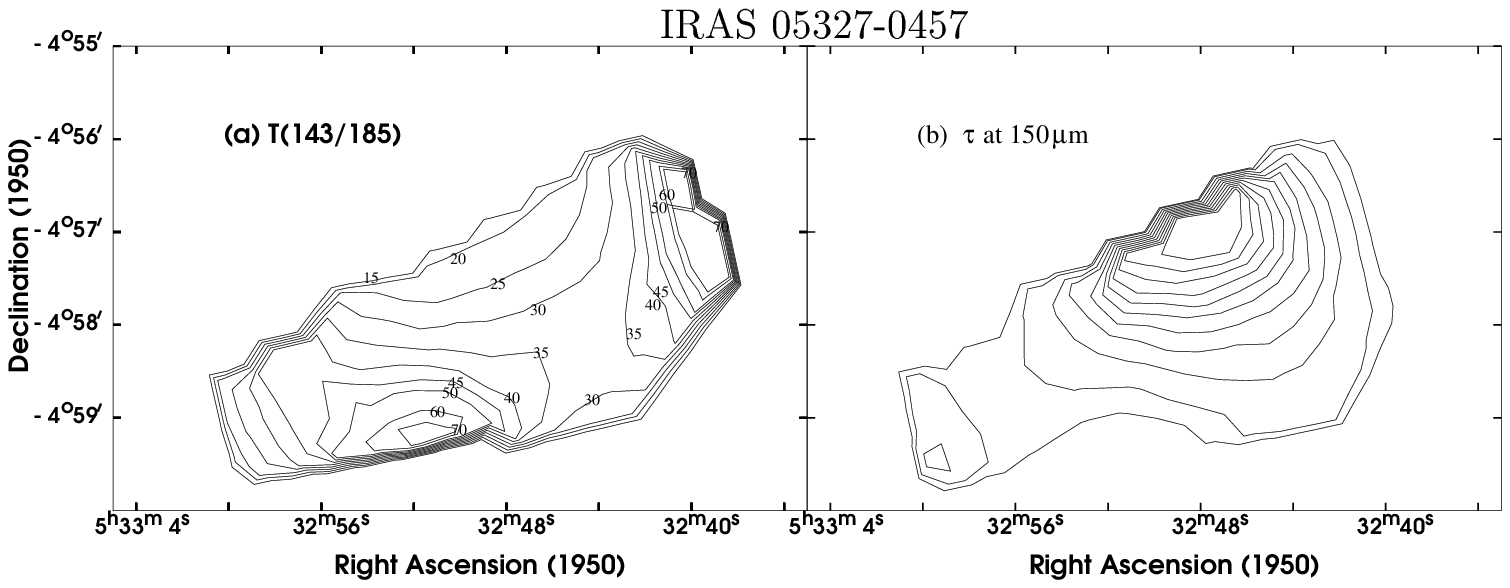]
{
(a) The dust temperature $T(143/185)$ and (b) optical depth \dep150\ maps
respectively for the region near IRAS 05327-0457 for
$\epsilon_{\lambda}\propto\lambda^{-1}$ emissivity law. The contours in
(a) are at 70, 60, 50, 45, 40, 35, 30,  25, 20 \& 15  K. The contours in
(b) as percentages of the peak value of 0.02 are same as the top 10 levels
of Fig. 2(b).
\label{ngc1977tmptau} 
}


\figcaption[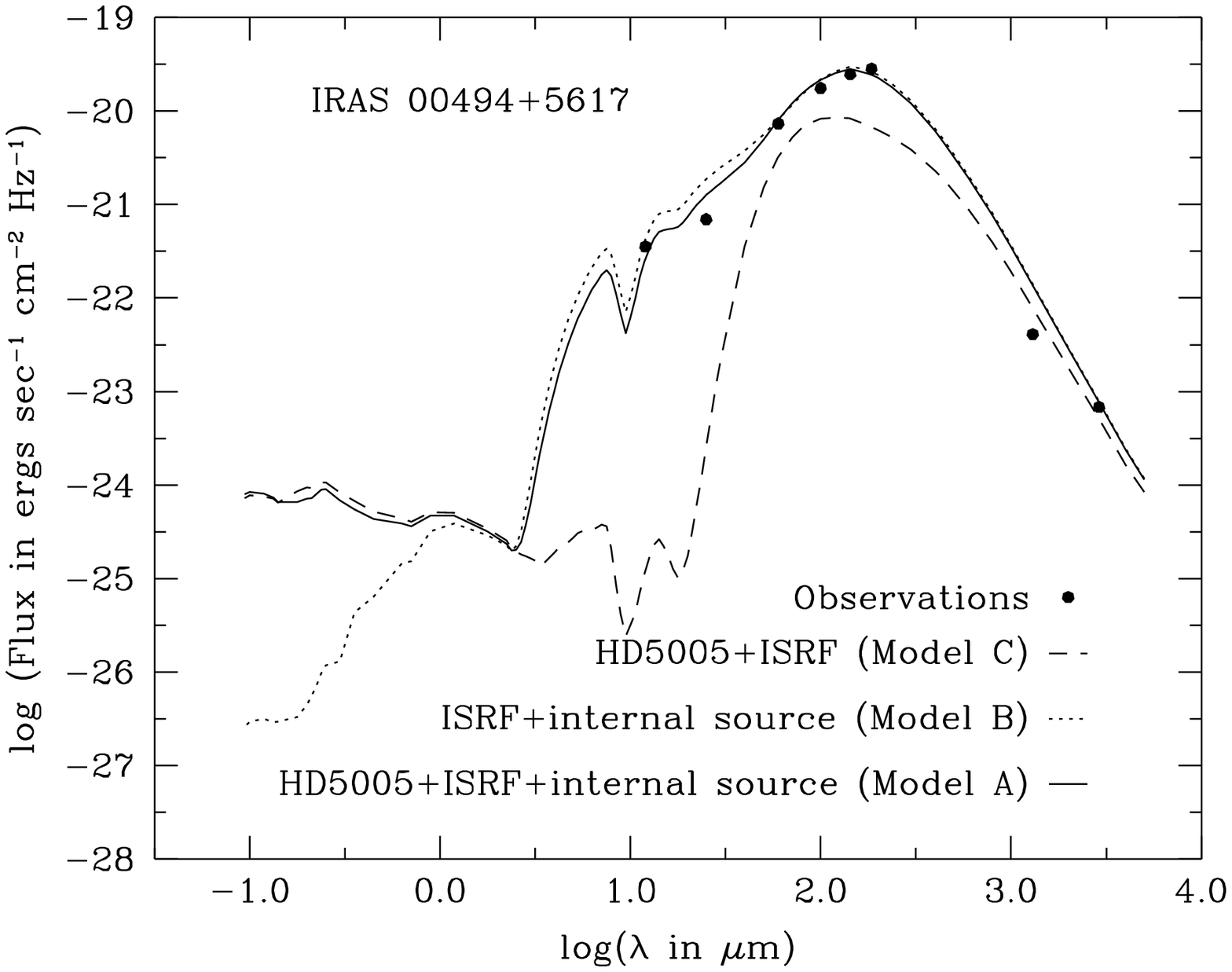]
{
Comparison of the observed spectral energy distribution with the predictions
from the models explored (see text) for IRAS 00494+5617. $A$ is the best fit
model.
\label{ngc281sed}
}

\figcaption[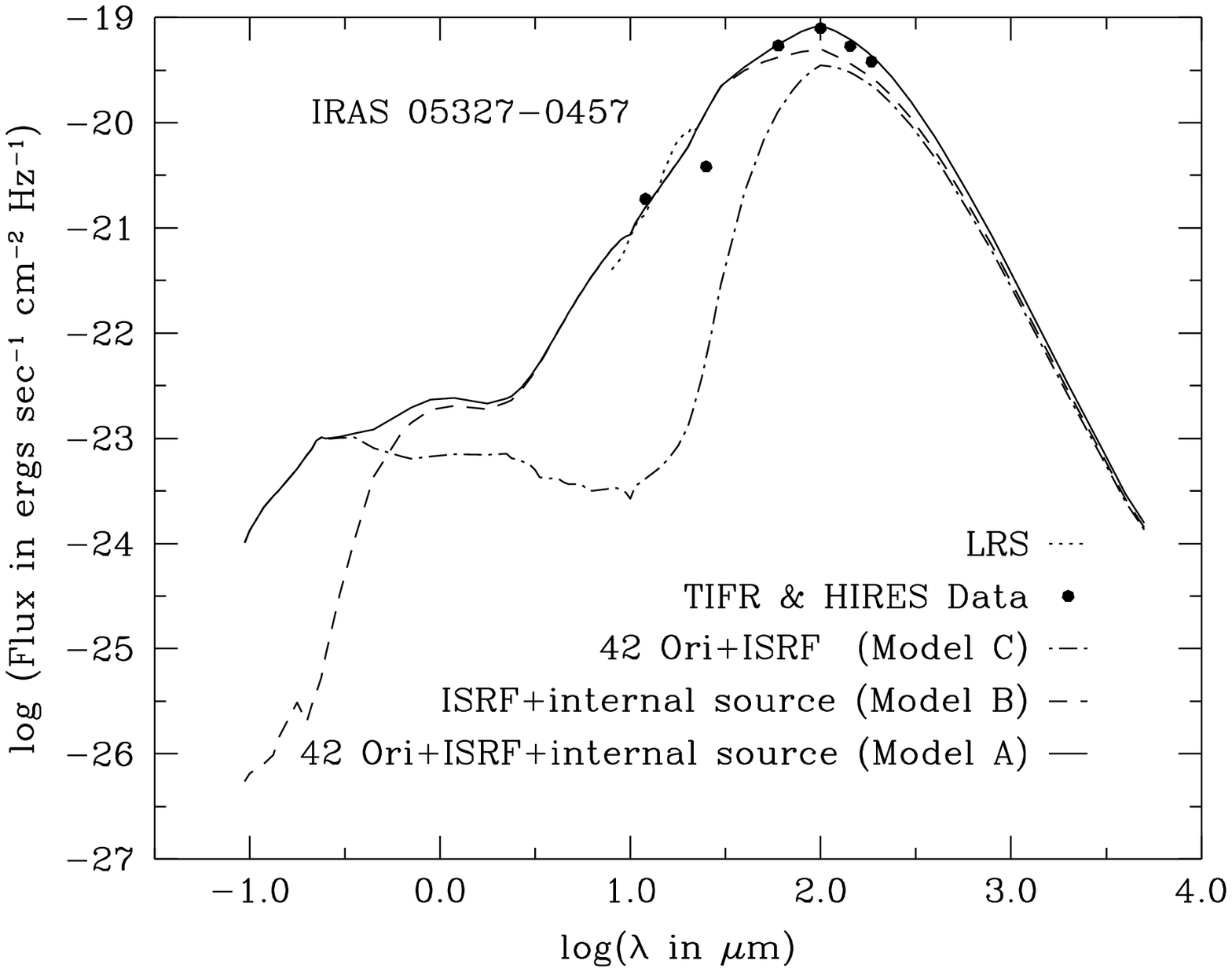]
{
Comparison of the observed spectral energy distribution with the predictions
from the models explored (see text) for IRAS 05327-0457. $A$ is the best fit
model. \label{ngc1977sed}
}



\begin{deluxetable}{lclll}
\footnotesize
\tablecaption{Positions and flux densities of the detected sources in the 
region around IRAS 00494+5617
\label{ngc281ftab}}
\tablewidth{0pt}
\tablehead{
\multicolumn{1}{l}{Source} &
\multicolumn{1}{c}{$\lambda$} &
\multicolumn{2}{c}{Position} &
\multicolumn{1}{c}{Flux }\\
\colhead{} &
\colhead{\micron\ } &
\colhead {$\alpha_{\rm 1950}$ } &
\colhead {$\delta_{\rm 1950}$}&
\colhead {Density} \\
& & \phn \phn 0$^{\rm h}$ & +56\arcdeg\ & \phn \phn Jy\tablenotemark{a}
} 
\startdata
           & 12  & 49$^{\rm m}$ 24\fs8 & 17\arcmin\ 42\arcsec\ & \phn \phn \phn 35 \\
           & 25  & 49 \phn 28.4 &  17\phn 27   &\phn \phn \phn 70 \\
           & 60  & 49 \phn 30.2 &  17\phn 42   &\phn \phn 733 \\
S 1        & 100 & 49 \phn 26.6 &  17\phn 27   &\phn 1762 \\
           & 143 & 49 \phn 29.9 &  17\phn 22   &\phn 2460 \\
           & 185 & 49 \phn 27.8 &  16\phn 31   &\phn 2820 \\
\hline
S 2        & 143 & 48 \phn 44.6 &  16\phn 39  &\phn\phn  115{\tablenotemark{b}} \\
           &  60 &  \phn \phn \phn "   &    \phn \phn "   &\phn  $<$25{\tablenotemark{b}} \\
           & 100 &   \phn  \phn \phn "  &   \phn  \phn "   & \phn $<$60{\tablenotemark{b}} \\
           & 185 &    \phn \phn \phn " &   \phn  \phn "   & $<$110{\tablenotemark{b}} \\
\enddata
\tablenotetext{a}{Flux density in 5\arcmin\ diameter circle  unless otherwise stated}
\tablenotetext{b}{Flux density in 2\arcmin\ diameter circle } 
\end{deluxetable}


\begin{deluxetable}{cclllc}
\footnotesize
\tablecaption{Positions and flux densities of the detected sources in the
region around IRAS 05327-0457
\label{ngc1977ftab}}
\tablewidth{0pt}
\tablehead{
\multicolumn{1}{l}{Source} &
\multicolumn{1}{l}{$\lambda$} &
\multicolumn{2}{c}{Position} &
\multicolumn{1}{c}{Flux }&
\multicolumn{1}{c}{Comment}\\
& \colhead{\micron} &
\colhead {$\alpha_{\rm 1950}$ } &
\colhead {$\delta_{\rm 1950}$}&
\colhead {Density} & \\
& & 5$^{\rm h}$ 32$^{\rm m}$& &\phn \phn Jy\tablenotemark{a} &  
}
\startdata
           & 12  &  46\fs9 & - 4\arcdeg\ 57\arcmin\ 28\arcsec\  &\phn \phn 191&  \\
           & 25  &  45.9 & - 4 \phn 57 43   &\phn \phn 383  &  IRAS \\
P 1        & 60  &  45.9 & - 4 \phn 56 58   &\phn 5401 & 05327-0457  \\
           & 100 &  45.9 & - 4 \phn 56 58   &\phn 7896 &   \\
           & 143 &  45.4 & - 4 \phn 57 47   &\phn 5350 &   \\
           & 185 &  46.4 & - 4 \phn 57 38   &\phn 3810 &   \\
\hline
P 2        & 143 &  56\fs6 & - 5 \phn 03 50   & \phn \phn 580\tablenotemark{b} & MMS 6  \\
           & 185 &  57\fs0 & - 5 \phn 03 25   &\phn 1820\tablenotemark{c} &   \\
\enddata
\tablenotetext{a}{Flux density in 5\arcmin\ diameter circle unless otherwise stated}
\tablenotetext{b}{Flux density in 2\arcmin\ diameter circle} 
\tablenotetext{c}{Flux density in 3\arcmin\ diameter circle}
\end{deluxetable}


\begin{deluxetable}{cccc}
\footnotesize
\tablecaption{Comparison of model predictions of FWHMs with observations.
\label{proftab}}
\tablewidth{0pt}
\tablehead{
\multicolumn{1}{c}{Source} &
\multicolumn{1}{c}{$\lambda$}  &
\multicolumn{1}{c}{Model} &
\multicolumn{1}{c}{Observations}\\
\colhead{} &
\colhead{\micron} &
\colhead{} &
\colhead{(major$\times$minor)} 
}
\startdata
            & 25  & 0\farcm6 & 1\farcm8$\times$1\farcm6 \\
            & 60  & 1\farcm0 & 2\farcm6$\times$1\farcm8 \\
IRAS 00494+5617  & 100 & 2\farcm2 & 2\farcm3$\times$2\farcm0 \\
            & 143 & 2\farcm3 & 3\farcm4$\times$2\farcm3 \\
            & 185 & 2\farcm5 & 2\farcm3$\times$1\farcm8 \\
\hline
            & 60  & 1\farcm5 & 5\farcm5$\times$2\farcm2 \\
IRAS 05327-0457  & 100 & 2\farcm8 & 7\farcm3$\times$3\farcm1 \\
            & 143 & 2\farcm7 & 5\farcm0$\times$1\farcm8 \\
            & 185 & 3\farcm0 & 3\farcm2$\times$1\farcm9 \\
\enddata
\end{deluxetable}


\begin{deluxetable}{cccccccccccc}
\footnotesize
\tablewidth{0pt}
\tablecaption{Parameters for the best fit radiation transfer model
\label{tranparm}}
\tablehead{
\multicolumn{1}{c}{Source} &
\multicolumn{1}{c}{$\beta$} &
\multicolumn{1}{c}{R$_{\rm max}$} &
\multicolumn{1}{c}{R$_{\rm min}$ }&
\multicolumn{1}{c}{Dust type }&
\multicolumn{1}{c}{ $\tau_{\rm 100}$ }&
\multicolumn{1}{c}{Luminosity} &
\multicolumn{1}{c}{Dust Composition }\\
&& & & & & & \colhead{Silicate:Graphite}\\
& & \colhead{(pc)} &
\colhead{(pc)} & &&
\colhead{(10$^{3}$ \lsun\ )}&
\colhead{\% : \%} 
}
\startdata
IRAS 00494+5617 & 0.0 & 1.2\phn\phn & 0.0001 & DL\tablenotemark{a}  & 0.1$\pm$0.01& 12$\pm$2 & 32:68  \\
IRAS 05327-0457 & 0.0 & 0.35\phn & 0.0001 & MMP\tablenotemark{b} & 0.03$\pm$0.005&  2.8$\pm$0.2 & 11:89  \\
\enddata
\tablenotetext{a}{DL \-- Draine \& Lee (1984).}
\tablenotetext{b}{MMP \-- Mathis, Mezger \& Panagia (1983).}
\end{deluxetable}

\end{document}